\newif\ifuseprd
\newif\ifom
\useprdfalse 
\omfalse    

\newif\ifatitp
\atitpfalse 

\ifuseprd 
\documentclass[preprint,tightenlines,floats,prd,eqsecnum,nobibnotes%
\ifatitp,superscriptaddress\fi,nofootinbib]{revtex4}
\else
\documentclass[final,12pt,letterpaper]{JHEP}
\fi 
\usepackage{amsmath,amssymb}

\newif\ifspinpm 
\spinpmtrue 

\allowdisplaybreaks[3]
\ifom
\input dnmacs
\fi 

\def\omt{{\ifom{{\dn\dnhalf :}}\else%
        {{3\!{\footnotesize$\mathbf{{\frown}\llap{\text{\tiny$\prime$}}}$}%
        {\hbox to -.7ex{\null}\llap{\raise1.3ex\hbox{\tiny{%
        \setbox255=\hbox{$\mathbf{{\smile}}$}%
        \copy255\kern-.7\wd255{\raise.5ex\hbox{$\mathbf{\cdot}$}}}}}}}}\fi}}

\newcommand\comment[1]{{\em [{#1}]}}
\newcommand\skipthis[1]{{}}

\makeatletter
\def\@strike{\relax\leavevmode
  \ifmmode
    \expandafter\mathpalette\expandafter\math@strike
  \else
    \expandafter\make@strike
  \fi}
\def\math@strike#1#2{%
  \setbox\z@\hbox{$\m@th#1{#2}$}\fin@strike}
\def\make@strike#1{%
  \setbox\z@\hbox{\color@begingroup#1\color@endgroup}\fin@strike}
\def\fin@strike{%
  \@tempdima\dp\z@
  \@tempdimb\ht\z@
  \lower\@tempdima\hbox{\strike@start}%
  \box\z@
  \raise\@tempdimb\hbox{\strike@end}}
\def\strike@start{\special{ps: %
    currentpoint /starty exch def /startx exch def}}
\def\strike@end{
}
\newcommand\fs{\protect\@strike}

\let\oldAE\AE
\renewcommand\AE{{\ifmmode{\text{\it\oldAE}}\else{\oldAE}\fi}}

\newcommand\ct[1]{{\ifuseprd{\em{#1}},\else{\sf {#1}},\fi}}
\newcommand\bt[1]{{\em {#1}},}
\newcommand\web[1]{{\tt \hbox{{#1}}}}
\newcommand\phepth[1]{{\tt [\hepth{#1}]}}

\DeclareMathOperator{\ch}{ch}

\chardef\til=`~

\newcommand\half{{\ensuremath{\frac{1}{2}}}}
\newcommand\p{\ensuremath{\partial}}

\newcommand\order[1]{{\ensuremath{{\mathcal O}({#1})}}}

\newcommand\anti[2]{\ensuremath{\left\{{#1},{#2}\right\}}}
\newcommand\com[2]{\ensuremath{\left[{#1},{#2}\right]}}
\DeclareMathOperator{\Tr}{Tr}

\newcommand\apr{{\ensuremath{{\alpha'}}}}

\newcommand\ep{\epsilon}

\newcommand\del{\partial}

\newcommand\ha{{\half}}

\newcommand\hA{{\hat{A}}}
\newcommand\hp{{\hat{\phi}}}
\newcommand\hF{{\hat{F}}}

\newcommand\cC{{\ensuremath{{\mathcal{C}}}}}
\newcommand\cX{{\ensuremath{{\mathcal{X}}}}}
\newcommand\co[1]{{\ensuremath{{\iota_{{#1}}}}}}
\newcommand\Pb{{\ensuremath{{\mathcal{P}}}}}

\DeclareMathOperator{\Pf}{Pf}
\DeclareMathOperator{\STr}{STr}

\newif\ifwz
\wzfalse
\newif\ifrr
\rrfalse
\newcommand\WZ{{\ifwz{WZ}\else{Wess-Zumino (WZ)}\fi\global\wztrue}}
\newcommand\RR{{\ifrr{RR}\else{Ramond-Ramond (RR)}\fi\global\rrtrue}}
\newcommand\swz{{\ensuremath{S_{\text{WZ}}}}}

\providecommand\putabstract[1]{\ifuseprd\begin{abstract} {#1} \end{abstract}%
                           \else \abstract{{#1}} \fi}
\providecommand\plb[3]{{Phys.\ Lett.\ B {\bf {#1}}, {#3} ({#2})}}
\providecommand\npb[3]{{Nucl.\ Phys.\ {\bf B{#1}}, {#3} ({#2})}}
\providecommand\jhep[3]{{J.\ High Energy Phys.\ {\bf #1}, {#3} ({#2})}}

\providecommand\ptp[3]{{\ifuseprd\else\begingroup\em\fi Prog.\ Theor\ Phys.\ %
     \ifuseprd\else\endgroup\fi {\bf {#1}}\ifuseprd, {#3} ({#2})\else%
     \ ({#2}) {#3}\fi}}
\providecommand\cqg[3]{{\ifuseprd\else\begingroup\em\fi Class.\ and Quant.\ %
     Grav.\ %
     \ifuseprd\else\endgroup\fi {\bf {#1}}\ifuseprd, {#3} ({#2})\else%
     \ ({#2}) {#3}\fi}}
\providecommand\cjm[3]{{\ifuseprd\else\begingroup\em\fi Canad.\ J.\ Math.\ %
     \ifuseprd\else\endgroup\fi {\bf {#1}}\ifuseprd, {#3} ({#2})\else%
     \ ({#2}) {#3}\fi}}
\newcommand\citeprd[3]{{\ifuseprd{Phys.\ Rev.\ D {\bf {#1}}, {#3} ({#2})}%
                        \else{\prd{#1}{#2}{#3}}\fi}}
\providecommand\hepth[1]{{\tt hep-th/{#1}}}

\ifuseprd
\begin{document} 
\fi 

\title{Ramond-Ramond Couplings of Noncommutative D-branes}
\ifuseprd
\author{Hong Liu}\email{liu@physics.rutgers.edu}
\affiliation{New High Energy Theory Center,
       Rutgers University;
       126 Frelinghuysen Road;
       Piscataway, NJ \ 08854}
\author{Jeremy Michelson}\email{jeremy@physics.rutgers.edu}
\affiliation{New High Energy Theory Center,
       Rutgers University;
       126 Frelinghuysen Road;
       Piscataway, NJ \ 08854}
\ifatitp
\affiliation{Institute for Theoretical Physics,
University of California;
Santa Barbara, CA \ 93106-4030}
\fi
\else 
\author{Hong Liu\ifatitp$^1$\fi\thanks{\tt liu@physics.rutgers.edu} \ 
and
Jeremy Michelson\ifatitp$^{1,2}$\fi\thanks{\tt jeremy@physics.rutgers.edu} \\
{\ifatitp$^1$\fi}New High Energy Theory Center \\
Rutgers University \\
126 Frelinghuysen Road \\
Piscataway, NJ \ 08854 \ USA 
\ifatitp\\
\vspace{\baselineskip}
$^2$Institute for Theoretical Physics \\
University of California \\
Santa Barbara, CA \ 93106-4030
\fi} 
\fi 

\putabstract{
We propose the leading couplings, in an $\apr$ expansion,
of 
D-branes to \RR\ potentials in a constant NSNS $B$-field for an
arbitrary choice of noncommutative parameter. 
The proposal is motivated by some
string amplitude computations.
The zero momentum 
couplings are topological in 
nature and include Elliott's formula involving the 
noncommutative Chern character. The finite
momentum couplings are given by smearing the zero momentum operators 
along an open Wilson line. Comparisons between the \RR\ couplings in 
different descriptions lead to a
better understanding of the field redefinitions between gauge field 
variables (the Seiberg-Witten map) and help constrain $\apr$ corrections. 
In particular we recover the Seiberg-Witten map conjectured by one of 
the authors in \hepth{0011125}. We also discuss the dynamics of the 
transverse scalar fields and find evidence for
a new derivative-driven
dielectric effect.
}

\preprint{RUNHETC-2001-12\ifuseprd,~\else\\\fi 
          NSF-ITP-01-29\ifuseprd,~\else\\\fi {\tt hep-th/0104139}}

\ifuseprd
\maketitle
\else
\begin{document}
\fi 

\wzfalse\rrfalse

\section{Introduction} \label{sec:intro}

In recent years, various interesting insights have been obtained by 
considering D-branes in the presence of a background  Neveu-Schwarz 
$B$-field~\cite{cds,dh,sw}. While a constant $B$-field background 
can be gauged away in the noncompact spacetime in the closed string 
sector, it generates a rather nontrivial effect on the open strings 
on the D-branes: the low energy world-volume theory becomes noncommutative. 
The low energy theory on the D-branes is given by the Dirac-Born-Infeld 
action plus \WZ\ terms. In the past, great insights have been gained 
into issues like gauge theory dynamics, black holes and the AdS/CFT 
correspondence by studying the couplings between the D-branes and bulk 
closed string modes. In addition studying \RR\ couplings has also yielded 
the ``branes-within-branes'' phenomenon~\cite{bwbw,bwbd}, K-theory 
descriptions of D-branes~\cite{mm,witk} and the Myers dielectric 
effect~\cite{myers}.

In this paper, we continue the study of couplings of the
noncommutative D-branes
to the spacetime gravity fields, following~\cite{liu,lm3} (see 
also~\cite{dastrivedi,ooguri,ooguri1,mukhi}). 
In~\cite{liu,lm3}, we examined the Born-Infeld couplings
to the fluctuations of the closed string metric $g_{\mu\nu}$, dilaton 
and the $B$-field.  In this paper we will examine the couplings 
to \RR\ potentials.  

As pointed out in~\cite{sw}, there are different descriptions 
of the D-brane dynamics parameterized by a noncommutative parameter 
$\theta$. Different descriptions are related to one another by field 
redefinitions which preserve the gauge orbit. 
\skipthis{
For example, 
let $\hA$ denote gauge field variables in a description with a 
noncommutative parameter $\theta$ and  $A$ the gauge field variables 
in the commutative description. Then $\hA_{\mu} = \hA_{\mu} (\theta, A)$,
and the gauge transformations of $\hA$ map to those 
of $A$. 
}%
The open string parameters, i.e.\ the metric $G$, a two-form 
background $\Phi$ and the coupling constant $G_s$, associated with
each $\theta$-description
can be found from the closed string background%
\footnote{We shall keep the closed string background $g,B$ fixed 
and choose either $\Phi$ or $\theta$ as a free parameter to specify 
the description. 
}%
~\cite{sw}
\begin{equation} \label{swp}
\begin{split}
& \frac{1}{g + B} = \frac{1}{G + \Phi} + \frac{\theta}{2 \pi \apr}   \\
& G_s = g_s \left(\frac{\det (G+ \Phi)}{\det (g+B)} \right)^{\ha}
= g_s \left(\frac{\det G}{\det g} \right)^{\frac{1}{4}} 
\end{split}
\end{equation}
There are three cases of~\eqref{swp} which are of particular interests:
\begin{enumerate}
\item $\theta = 0$, $\Phi = B$. This is the commutative description.

\item $\Phi = 0$, $\frac{\theta}{2 \pi \apr}
= - \frac{1}{g+B} B \frac{1}{g-B}$. 
This is the description that most naturally follows from the on-shell 
string amplitudes. 

\item $\Phi = -B$, $\frac{\theta}{2 \pi \apr} =  \frac{1}{B}$. 
This is the Matrix model 
description~\cite{seiberg} and is closely related background 
independence of noncommutative Yang-Mills theory~\cite{sw,seiberg}.
\end{enumerate}

The leading order \RR\ couplings in the commutative description are well 
known~\cite{li,bwbd,myers} 
They are given by the \WZ\ terms%
\footnote{
$\mu_{p}$ is the \RR\ charge of the brane.
The pullback $\Pb$ and the
contraction $\co{\phi}$ of the transverse scalars
are defined in more detail in
section~\ref{results}, particularly equations~\eqref{defpb} and~\eqref{defco}.}
plus derivative corrections which are accompanied 
by  higher powers of $\apr$
\begin{equation} \label{dbc}
\swz = \mu_{p}
\STr \int \left(\Pb e^{- i\co{\phi}\co{\phi}} e^{B} C\right)e^F + \order{\apr} \ .
\end{equation}
In this paper we are interested in finding the leading order couplings  
(in an $\apr$ expansion) in a general $\theta$-description%
\footnote{The leading 
action for the $\Phi = -B$ description was proposed in~\cite{mukhi},
from the connection between noncommutative D-branes and the Matrix model. 
\skipthis{\begin{equation}
X^{\mu} = x^{\mu} + \theta^{\mu \nu} \hA_{\nu}, \quad
X^{i} = X^{i} (x^{\mu}), \quad [x^{\mu}, x^{\nu} ] = i \theta^{\mu \nu} \ .
\end{equation}
}}. 
Since the relations~\eqref{swp} between the open string parameters 
for different $\theta$-descriptions involve $\apr$, the effective action
in one description (e.g.\ with some $\theta_1$) should be considered as a 
resummation (along with field definitions between gauge field
variables) of that in another description (e.g.\ with some $\theta_2$)  
including all higher order $\apr$ corrections. For this reason, understanding 
the relations between different descriptions should be very useful for 
a deeper understanding of noncommutative gauge theories, and more generally,
D-brane dynamics.

We have computed  various tree-level disk amplitudes for the
scattering of open string gauge fields off an \RR\ potential, and extracted
the effective action to the lowest order in $\apr$. The \RR\ couplings 
found in this way correspond to the $\Phi=0$ description. There are several 
advantages to using this description.  Since it follows directly 
from on-shell amplitudes, the low energy gauge theory dynamics is most 
transparent in this description.%
\footnote{Recall that the mass shell conditions for massive 
open string states are defined in terms of the open string metric $G$
in the $\Phi =0$ description. For this reason, the leading order results
(in terms of $\apr$) for other values of $\Phi$ generally involve the 
contributions of massive open string modes.} \ 
Also varying $B$, interpolates between 
the commutative description, \hbox{$\Phi = B$}, and the Matrix 
description, \hbox{$\Phi = -B$}.
More explicitly, since (for \hbox{$\Phi=0$}), 
\hbox{$\theta= - (2 \pi \apr)  \frac{1}{g+B} B \frac{1}{g-B}$},
as \hbox{$B \rightarrow 0$}, $\theta \rightarrow 0$,  giving the 
commutative description, and as  \hbox{$B \rightarrow \infty$},
\hbox{$\theta \rightarrow (2 \pi \apr) \frac{1}{B}$},
which is the Matrix model description.
By contrast, in the \hbox{$\Phi = -B$} description, 
the \hbox{$B \rightarrow 0$} limit is the limit of infinite, not zero
noncommutativity.  Thus it is not as convenient a
description of the low energy
theory as the commutative gauge theory. 

We shall argue that the couplings we find for $\Phi =0$ actually
give the leading \RR\ couplings for all $\Phi$ descriptions. 
This is partly due to the topological nature of the leading couplings 
of the \RR\ potentials. 
The conclusion is also consistent with the known 
results for $\Phi = B$ and $\Phi = -B$.

Comparisons between the \RR\ couplings in different descriptions lead
to a
better understanding of the field redefinitions between gauge field 
variables (the Seiberg-Witten map).  They also constrain 
$\apr$ corrections. In particular we shall recover the Seiberg-Witten 
map conjectured by one of the authors in~\cite{liu}. We also derive 
some other interesting identities regarding the map.

While the amplitude calculations are in principle staightforward to carry out, 
the intermedate steps  turn out to be quite messy and complicated. 
In order to not have complicated 
intermediate technical details obscure the simplicity of the 
final results, and the physics therein,
we have separated this work into two parts, with this part containing the major
results and physical implications, and the other containing mostly 
detailed calculations.

This paper is organized as follows. In 
section~\ref{results} we present our results from amplitude calculations.
We give the \RR\ couplings in various equivalent forms, which are useful
for different purposes. In particular, we connect
the commutative result~\eqref{dbc} to the proposal of~\cite{mukhi} for
$\Phi = -B$.
In section~\ref{ktheory}, we discuss the  
relations of our results in section~\ref{results} with 
some results in noncommutative geometry, and their consistency with 
T-duality.%
\footnote{For the $\Phi=-B$ proposal of~\cite{mukhi}, T-duality was
discussed in~\cite{td}.} \ 
In section~\ref{phi} we compare the \RR\ couplings of different descriptions,
from which we confirm the Seiberg-Witten map conjectured in~\cite{liu} 
and make some remarks regarding 
$\apr$ corrections.
In section~\ref{transverse} we 
focus on the transverse scalar couplings and discuss the analogue of 
the 
Myers dielectric effect~\cite{myers}.
We conclude in section~\ref{conclusion} by discussing the possible 
relation between our discussion and anomalies in noncommutative 
gauge theories, and  offering some 
speculations for the couplings for non-BPS branes. 

As we were completing this work,~\cite{oo,mukhi2} appeared which has
some overlap with the contents of this paper, particularly section~\ref{phi}.

\section{The Wess-Zumino Action for Noncommutative D-branes} \label{results}

We have extrapolated the \RR\ couplings of a collection of
noncommutative D-branes, by computing tree-level disk diagrams for the
scattering of open string gauge fields off an \RR\ potential, and extracting 
the effective action.  The details are presented in~\cite{lm5}. The 
action found in
this way corresponds to the description with $\frac{\theta}{2 \pi \apr}
= - \frac{1}{g+B} B \frac{1}{g-B}$. Eventually (in section~\ref{phi}) 
we shall argue that it applies to any $\theta$-description.

We shall present the couplings in three equivalent ways which are useful 
for different purposes. We will discuss their physical implications 
in later sections. To better understand  the structure of the action, 
we shall first give the couplings for \RR\ potentials at zero momentum.
This captures the Lorentz tensor structure of the Yang-Mills operators. 
The finite momentum couplings turn out, not surprisingly,
to have the form of the zero momentum
operators, after smearing them along an 
open Wilson line~\cite{iikk,ambjorn,dasrey,ghi} following the
prescription given in~\cite{liu,dastrivedi}.

In the following, we shall use  $\hA_{\mu}$, $\hp^i$ 
to denote the noncommutative 
gauge fields and the transverse scalars.
Indices longitudinal to the brane are $\mu,\nu,\cdots$ 
and transverse indices are $i,j,\cdots$. 
When appropriate
we will assume that
$\theta$ and $B$ have maximal rank. Then, for $p$ odd, we shall consider a
Euclidean world-volume with all longitudinal directions noncommutative, while 
for $p$ even all longitudinal directions but time are noncommutative.

\subsection{Zero momentum couplings}

Our first expression has a close resemblance to 
the commutative couplings~\eqref{dbc}. At zero momentum, the 
\RR\ potentials $C=\sum_n C^{(n)}$  are constant forms and the 
couplings of $N$ noncommutative $Dp$-branes can be written as%
\footnote{Strictly speaking, the following formulas are only 
appropriate for topologically trivial configurations. See also
footnote~\ref{trace} in section~\ref{ktheory}.} 
\begin{equation} \label{rrf}
\swz = \mu_{p}\,  \STr \int  \left(e^{-\co{\tilde\theta}} e^{2\pi \apr \hat{F}} 
      \Pb\right)
  e^{-2 \pi \apr i  \co{\com{\hp}{\hp}}} \,  C  e^B,
\end{equation}
where $\STr$ is the symmetrized trace over the $U(N)$ matrices, 
$\mu_{p} = \frac{1}{(2 \pi)^p g_s \apr^{(p+1)/2}}$ 
is the D-brane charge ($g_s$ is 
the closed string coupling), and $\tilde{\theta}=\frac{\theta}{2\pi\apr}$.
As usual, wedge products are 
implied in expanding the exponential and in
products of forms, and the integration extracts only the $(p+1)$-form 
in the integrand.  In evaluating products of open-string fields, the
$\ast$-product is used.

The notation $\co{T}$ denotes contraction with respect to
the antisymmetric tensor $T$ of rank $m$; i.e.\
\begin{equation} \label{defco}
\left(\co{T} \omega^{(n)}\right)_{\mu_1 \dots \mu_{n-m}}
= \frac{1}{m!} T^{\nu_m \cdots \nu_1 } \omega_{\nu_1 \cdots \nu_m \mu_1 
\dots \mu_{n-m}}.
\end{equation}
The notation $\Pb$ denotes the pullback;
e.g.
\begin{equation} \label{defpb}
\Pb \omega^{(2)}_{\mu\nu} = \omega^{(2)}_{\mu\nu}
+ 2 (D_{[\nu} \hp^i) \omega^{(2)}_{\mu]i} 
+ D_{\mu} \hp^i D_{\nu} \hp^j \omega^{(2)}_{ij},
\end{equation}
where 
\begin{equation} \label{DX}
D_\mu \hp^i = \p_\mu \hp^i - i [\hA_{\mu}, \hp^i]
\end{equation}
and the square
brackets in~\eqref{defpb} denote antisymmetrization with unit weight.
Note that with a mild notational abuse,
\begin{equation} \label{altPb}
\Pb = e^{D\co{\hp}}.
\end{equation}
where $D\co{\hp}= D_{\mu} \hp^i d x^{\mu}$ is considered 
as a one-form in the worldvolume and a contracted vector in the transverse 
dimensions.  That is, we can think of $D\co{\hp}$ as an
operator which acts on forms to the right, by contracting the vector
index and antisymmetrizing the form index, thereby preserving the
dimension of the form on which it acts.  This reproduces~\eqref{defpb}.

The noncommutative field strength is 
\begin{equation} \label{ncf}
\hF_{\mu \nu} = \del_{\mu} \hA_{\nu} -  \del_{\nu} \hA_{\mu}
 - i [\hA_{\mu}, \hA_{\nu}],
\end{equation}
and $\hp^i$ are the transverse scalars.
Note that the commutators in~\eqref{DX},~\eqref{ncf} and in \co{\com{\hp}{\hp}}
in~\eqref{rrf} include both the $U(N)$ and $\ast$-products. 
In~\eqref{rrf}, the symmetrized trace treats gauge covariant objects 
$\hF$, $D_\mu \hp^i$ and $[\hp^i, \hp^j]$ as single operators, and all 
products between them are $\ast$-products.
The parentheses in~\eqref{rrf} enforce that $\tilde{\theta}$ can only contract
with $\hF$ and/or the $D_{\mu}$s from $\Pb$, but with nothing else.
For example, with $C^{(2)} = \left[\frac{1}{2}C_{\mu\nu} dx^{\mu} dx^{\nu}
      + C_{\mu i} dx^\mu dx^i
      + \frac{1}{2} C_{ij} dx^i dx^j \right]$,
\begin{multline} \label{ex(iP)C}
\left(e^{-\co{\theta}} \Pb\right) C^{(2)}
= \left[\frac{1}{2}C_{\mu\nu} dx^{\mu} dx^{\nu}
      + C_{\mu i} dx^\mu dx^i
      + \frac{1}{2} C_{ij} dx^i dx^j \right]
\\
+ D_{\nu} \hp^i \left[C_{\mu i} dx^\mu dx^\nu + C_{ij} dx^\nu dx^j
      \right]
+ D_{\mu} \hp^i D_{\nu} \hp^j C_{ij} dx^\mu dx^\nu
- \frac{1}{2} \theta^{\tau\sigma} D_{\sigma} \hp^i D_{\tau} \hp^j C_{ij}.
\end{multline}
For convenience,
in the following we shall set the inverse string tension $2 \pi \apr
=1$.

An alternative way of writing~\eqref{rrf} is 
\begin{equation} \label{asXX}
\swz = \frac{\mu_{p}}{\Pf(\theta)} \STr 
\int  \star 
\left(e^{-i \co{\com{X}{X}}} \; 
e^{B-\frac{1}{\theta}} C \right),
\end{equation}
where $\star$ is the Hodge dual%
\footnote{The metric dependence of the Hodge dual drops out in these
equations.}
in the {\em noncommutative}%
\footnote{That is, for odd $p$, we take the Hodge dual in the brane,
whereas for even $p$, we take the Hodge dual along the spatial
directions of the brane.
Thus, for odd $p$, equation~\eqref{asXX} is equivalent to
$\swz = \frac{\mu_p}{\Pf(\theta)} \STr 
\int  
d^{(p+1)} x \,  
\left(e^{-i \co{\com{X}{X}}} \; 
e^{B-\frac{1}{\theta}} C \right)$.
For even $p$, equation~\eqref{asXX} can be similarly rewritten as the
integral of a time-like one-form.
\label{ft:hd}}
directions and
\begin{equation}
X^{\mu} = x^{\mu} + \theta^{\mu \nu} \hA_{\nu}, \quad
X^{i} = \hp^{i} (x^{\mu}), \quad [x^{\mu}, x^{\nu} ] = i \theta^{\mu \nu} \ .
\end{equation}
Note that 
\begin{equation}
[X^{\mu}, X^{\nu} ] = i \left(\theta - \theta \hF \theta \right)^{\mu \nu},
\quad 
[X^{\mu} , X^i] = i \theta^{\mu \nu} D_{\nu} \hp^i \ .
\end{equation}
In reaching equation~\eqref{asXX}, and for the rest of this subsection,
we need to take $\theta$ to be of maximal rank, in the sense described at the end of
the introduction to this section.

While it is not obvious, the equivalence of~\eqref{asXX}
to~\eqref{rrf} can be shown by repeatedly using the 
formulas
\begin{gather} \label{moveco}
 \int \left(e^{-\co{\theta}} \chi\right) \omega 
= \int \chi \; e^{-\co{\theta}} \omega, \\
\label{itow}
e^{-\co{\theta}} \omega = \Pf(\theta) \star \left( \left(\star
e^{-\theta^{-1}} \right) (\star \omega) \right),
\end{gather}
and an identity 
\begin{equation} \label{cbh}
e^{D\co{\hp}} e^{-\co{\theta}} = e^{-\co{\theta}} e^{\frac{1}{\theta}}
  e^{\co{\theta D\hp}} e^{-\frac{1}{\theta}}.
\end{equation}
where in~\eqref{cbh} we have used the notation~\eqref{altPb} and 
$\theta  D\hp$ is a short-hand for $\theta^{\mu \nu} D_{\nu} \hp^i$.

A third way of writing~\eqref{rrf} is 
\begin{equation} \label{trr}
\begin{split}
\swz & = \mu_p \, \STr \int \left( e^{- \co{\theta}} e^{\hF} \right) \left(
e^{\frac{1}{\theta}} \, e^{\co{\theta D \hp}} \, e^{-i \co{\com{\hp}{\hp}}} \;
e^{B -\frac{1}{\theta}} \; C \right) \\
& = \mu_p  \STr \int \sqrt{\det (1 - \theta \hF)} \; 
e^{\hF \frac{1}{1- \theta \hF}}
\left( e^{\frac{1}{\theta}} \, e^{\co{\theta D \hp}} \, 
e^{-i \co{\com{\hp}{\hp}}} \; e^{B -\frac{1}{\theta}} \; C \right)
\end{split}
\end{equation}
which can again be derived from~\eqref{rrf} or~\eqref{asXX} 
by using~\eqref{moveco}--\eqref{cbh}. In the second line above we have 
used an identity 
\begin{equation} \label{ith} 
e^{- \co{\theta}} e^{\hF} =
\sqrt{\det (1 - \theta \hF)} \; e^{\hF \frac{1}{1- \theta \hF}}
\end{equation}
Note that
in contrast with~\eqref{rrf}, in~\eqref{trr} $e^{- \co{\theta}}$ acts
only on $e^{\hF}$. 

Since in the above discussion, $\theta
= - \frac{1}{g+B} B \frac{1}{g-B}$, if $B \ll g$ then
$\theta\sim B$, and if $B \gg g$, then
$\theta \sim \frac{1}{B}$.  Thus the action~\eqref{rrf} and/or~\eqref{asXX} 
interpolates between~\eqref{dbc} for $B=0$, to
\begin{equation} \label{matrix}
\swz = \frac{\mu_p}{\Pf(\theta)} \STr 
\int \star  \left(e^{-i \co{\com{X}{X}}} \;  C \right) \ ,
\end{equation}
for $B=\infty$.
Equation~\eqref{matrix} is precisely the proposal of~\cite{mukhi} 
based on the connection with the Matrix model. While~\eqref{asXX}
and~\eqref{matrix}
are tantalizingly close, they differ for general values of $B$ and are 
related by complicated 
field redefinitions and resummation over $\apr$ corrections.

It is interesting to note that if we keep $B$ fixed, and simply take 
$\theta = 0$ or  $\theta = \frac{1}{B}$%
\footnote{Of course, the $\hA$ and $\hp$  in~\eqref{rrf} and~\eqref{asXX}
are associated with $\theta = - \frac{1}{g+B} B \frac{1}{g-B}$. when
taking $\theta = \frac{1}{B}$ (or $0$) we also need to take $\hA$ and $\hp$ to 
be the variables assciated with $\theta = \frac{1}{B}$ (or $0$).}
we precisely recover~\eqref{dbc} or~\eqref{matrix} respectively. 
This observation will be 
important for our later claim that~\eqref{rrf} and~\eqref{asXX} apply 
to all $\theta$-descriptions in section~\ref{phi}.

\subsection{Finite momentum couplings} \label{fmc}

At finite momentum $q$, from the results in~\cite{liu,dastrivedi}, 
one expects the presence of a generalized open Wilson line
$W(x,\cC_q)$~\cite{lm3,ooguri} 
\begin{equation} \label{gowl}
W(x,\cC_q)  =  P_{\ast} \exp \left[ i 
\int_0^1 d \tau  \,
\left(q_{\mu} \theta^{\mu \nu} \hA_{\nu} (x + \xi (\tau))
+ q_{ i}  \hp^i ( x + \xi (\tau)) \right) \right].
\end{equation}
\skipthis{
W[\cC] = P_\ast
\exp\left[i \int_{\cC} (q_\mu \theta^{\mu\nu} A_\nu + q_i
\hp^i)\right]
}%
where $q = (q^{\mu}, q^i)$ is the spacetime momentum of the \RR\ potentials
and the contour $\cC_q$ is a straight-line 
$\xi (\tau) = \theta^{\mu\nu} q_{\nu} \tau$.
In addition the Yang-Mills operators which appear in the zero momentum 
couplings~\eqref{rrf},~\eqref{asXX},~\eqref{trr} should now be smeared 
along the Wilson line~\eqref{gowl}~\cite{liu,dastrivedi}. This 
prescription, which has been 
checked explicitly in the couplings of noncommutative D-branes to supergravity
fields in the NS-NS sector in~\cite{lm3,ooguri}, is  confirmed by our 
amplitude calculations in the \RR\ case as well.

More explicitly, the coupling of $N$ noncommutative 
$Dp$-branes to \RR\ fields can be written in momentum  space as%
\footnote{
Note that in~\eqref{rrct}, the  momentum integrations are over all 
directions, including the transverse components which are not conserved by the 
D-branes.  From the
D-brane point of view, this is reflected in its dependence on 
transverse coodinates through the factor  $e^{i q \cdot \hp}$
in the generalized Wilson line~\eqref{gowl}.
For example, in the case $\theta \rightarrow 0$, the contour of the Wilson 
line~\eqref{gowl} shrinks to a point and~\eqref{gowl} becomes $\exp (i
q_i \phi^i)$;
thus integration over $q^i$ yields 
$\int d q^i \, C(q) \, \exp (i q_i \phi^i) = C (\phi (x), q^{\mu})$.
}%
\begin{equation} \label{rrct}
\swz = \mu_p \, \int d^{10} q \; O_C (-q)\; C (q) e^B,
\end{equation}
where $O_C (-q)$ is a gauge invariant operator
constructed from  $O_C (x)$, the Yang-Mills operator appearing in the 
integrand 
of~\eqref{rrf} (or equivalently~\eqref{asXX} and~\eqref{trr}), and
$W(x, \cC_q)$ in~\eqref{gowl} via
\begin{equation} \label{fourier}
O_C (q) = \Tr \int d^{p+1} x \; L_{\ast} \left[ W(x,\cC_q) 
   O_C (\hA,\hp)\right] 
\ast e^{iq_{\mu} x^{\mu}},
\end{equation} 
where the notation $L_{\ast}$, introduced in~\cite{liu}, 
is a short-hand for the prescription in which each single operator 
(i.e. one of $\hF$ , $D_\mu \hp^i$,  $[\hp^i, \hp^j]$)  contained in 
$O_C (x)$ is integrated independently over the Wilson
line~\eqref{gowl} using the path ordering with respect to the 
$\ast$-products. For example, if we had $O_C (x) = \hF(x) \wedge   
\hF (x)$, then we would have
\begin{equation} \label{giop}
\begin{split} \raisetag{3\baselineskip}
O_C (q) & = \Tr \int d^{p+1} x \; L_{\ast} \left[ W(x,\cC_q) \hF(x) \wedge   
\hF (x) \right] \ast e^{iq_{\mu} x^{\mu}} \\
& =  \Tr \int d^{p+1} x  \! 
 \int^{1}_{0}\! d \tau_1 d \tau_2 
 \; P_{\ast} \left[ W(x,\cC_q) 
     \hF (x + \xi(\tau_1)) \wedge \hF (x + \xi(\tau_2))
 \right] \ast e^{i q _{\mu} x^{\mu}} 
\skipthis{
& = \sum_{m=0}^{\infty} \int  d^{p+1} x  \, Q_{m} (x) e^{iq _{\mu} x^{\mu}}
}
\end{split}
\end{equation}
On expanding the Wilson line and performing the $\tau$ integrations, $O_C (q)$ 
can be written in terms of a power series of $\hA$ and $\hp$ using 
$n$-ary operations $\ast_n$~\cite{garousi,lm2}.
We refer to~\cite{liu} for the definition and properties 
of the $n$-ary operations and their the relations to 
the expansion of open Wilson lines.
Note that the
$L_{\ast}$-prescription symmetrizes the integrand, and so
the symmetrized trace prescription in~\eqref{rrf} 
is recovered at zero momentum.

\section{T-duality, Elliot formula and K-theory} \label{ktheory}

In this section we shall be interested in the zero momentum couplings
of the gauge fields and set  the transverse scalar fields to zero, in which 
the case the action~\eqref{rrf} (or~\eqref{trr})  becomes%
\footnote{ \label{trace}
Strictly speaking, for topologically nontrivial configurations, 
there is not a meaningful separation of  the trace $\Tr$ over the group 
indices and the spacetime integration $\int$ in a noncommutative space.  
Instead one simply gets a $\Tr_{\theta}$ which combines both, and 
in the trivial case becomes $\Tr_{\theta} = \Tr \int$. 
Another caveat
is that when we take the zero momentum limit, the open Wilson line might 
give rise to
a nontrivial normalization factor, which is also incorprated in $\Tr_{\theta}$.
We would like to thank C. Hofman for  discussions regarding 
these issues.
}
\begin{equation} \label{grr}
\swz  = 
\mu_{p} \, \Tr_{\theta} \left( e^{- \co{\theta}} e^{\hF} \right)\; e^{B} C 
= 
\mu_{p} \, \Tr_{\theta} 
\sqrt{\det (1 - \theta \hF)} \; e^{\hF \frac{1}{1- \theta \hF}} \; 
e^{B} C 
\end{equation}
It may seem somewhat surprising 
that we  have an $e^B$ factor in the noncommutative description, since 
one might expect to have only 
open string parameters in the action. The reason is the following. Suppose we 
compactify the theory on a torus $T^d$. It is well known that the scalar 
fields from the \RR\ potentials transform in a chiral spinor representation 
of the T-duality group. However, as was argued in~\cite{bmz,fot}, 
the fields which transform
covariantly under the $SO(d,d;{\mathbb{Z}})$
group involve not only the R-R fields $C$ 
but also the Neveu-Schwarz
$B$-field in a combination  which is precisely $D= C e^B$.   
The presence of the factor $e^B$ also suggests that if we had included 
the fluctuations of the $B$-field, it might simply enter the 
noncommutative description as $B \rightarrow B + \delta B$ in~\eqref{rrf}. 
It would be interesting to check whether 
equation~\eqref{rrf} incorporates fluctuations of the
$B$-field correctly. We have not computed any amplitudes with 
both \RR\ and NS-NS vertex operators, 
and so our computation was not sensitive to fluctuations of $B$.
\skipthis{Of course, we know that in the commutative limit,~\eqref{grr} does
properly include fluctuations of $B$.}

The \RR\ couplings exhibit the  ``branes within branes'' 
phenomenon~\cite{bwbd}, 
where D-branes of lower dimensions are described by topologically
nontrivial configurations of the world-volume gauge theory. In the 
noncommutative case, the counterparts of gauge bundles are finite 
projective modules of the noncommutative algebra, which can be---modulo  
physical processes of brane-anti brane creation and 
annihilation---classified by the K-theory group of the algebra.
The charges of lower dimensional branes, from equation~\eqref{grr},
instead of being given by the Chern character 
$\ch (E) = \Tr_{\theta} e^{\hF}$ 
as in the commutative description~\eqref{dbc}, are given by
\begin{equation} \label{ch}
\mu (E) =e^{-\co{\theta}} \ch (E) 
 = \Tr_{\theta} e^{-\co{\theta}} e^{\hF} = 
\Tr_{\theta} \sqrt{\det (1 - \theta \hF)} \; e^{\hF \frac{1}{1- \theta \hF}}
\end{equation}
where $E$ above denotes the gauge bundle (or the projective module)  
on the brane. 

This fits very well with the results in noncommutative geometry and
the K-theoretic description of the D-branes. For a commutative manifold
${\cal M}$,
the Chern character maps the elements of the K-theory group $K^0 ({\cal M}) $ 
of the manifold  to the integral elements of the even cohomology 
class. However, in the noncommutative case---for concreteness let us 
consider the example of a noncommutative torus $T^d_{\theta}$---the Chern 
character $e^{\hF}$, while still defining a 
homomorphism from $K^0 ({\cal M}) $ to the elements of the cohomology class, 
does not  map to the integral elements (see e.g.\cite{cds,schwarz,hofman}). 
On the other hand  it was shown by Elliott~\cite{elliott,rieffel} that 
the K-group $K^0 (T^d_{\theta})$ can be identified with the 
even integral cohomology 
lattice of $T^d_{\theta}$ and in particular  a K-theory
class $\mu (E) $ of a module $E$ can be computed from its Chern character
precisely using~\eqref{ch} (see e.g.~\cite{schwarz,schwarzre}). 
Equation~\eqref{ch} is also 
called the Elliott formula.
\skipthis{
\footnote{The K-theory group of a noncommutative torus is the same 
as that for the commutative tori:$K_{0} (T^d_{\theta}) = Z^{2^{d-1}}$, whose 
elements transform under a chiral spinor representation of the T-duality
group $SO(d,d,Z)$}. 
}

\section{Relations between different descriptions and 
the Seiberg-Witten Map} \label{phi}

In this section we shall again set  the transverse scalar fields to zero
and be interested in the couplings of the gauge fields at finite 
momentum, in which case the action becomes
\begin{equation} \label{gfmc}
\swz = \mu_p \, \int d^{p+1} k \; Q(k) D (-k) 
\skipthis{
\int \left[Q_0 D^{(p+1)} + Q_2  D^{(p-1)}
+  Q_4  D^{(p-3)} + \cdots \right] \ ,
}
\end{equation} 
where $D = C e^{B}$ and 
\begin{equation} \label{gfmc1}
\begin{split}
Q(k) & =  \Tr \int d^{p+1} x \; L_{\ast} \left[e^{- \co{\theta}} e^{\hF}
\;  W(x,\cC_k) \right]\ast e^{ik \cdot x} \\
&  =  \Tr \int d^{p+1} x \; L_{\ast} \left[
\sqrt{\det(1-\theta \hF)} 
e^{\hF \frac{1}{1-\theta \hF}} \; W(x,\cC_k) \right] \ast e^{ik \cdot x} \\
& = \int d^{p+1} x \; Q(x)  \, e^{ik \cdot x} \ .
\end{split}
\end{equation}
In the third line we have defined $Q(x)$ as 
the fourier transform of $Q(k)$ to coordinate space.
Note that $Q(x)$ can not be simply identified with the integrand of the 
first two lines since the path of the Wilson line depends on $k$. 

We may expand $Q$ in terms of differential forms of different degrees, i.e. 
\begin{equation}
Q(k) = Q_0 (k) + Q_2 (k) + Q_{4} (k) + \cdots 
\end{equation}
with $Q_0, Q_2, Q_4$ a scalar, 2-form and 4-form respectively. For example, 
\begin{align}
Q_{0} (k) & =  \Tr \int d^{p+1} x \; L_{\ast} \left[
\sqrt{\det(1-\theta \hF)}\;  W(x,\cC_k) \right] \ast e^{ik \cdot x} \\
Q_{2} (k) & =  \Tr \int d^{p+1} x \; L_{\ast} \left[
\sqrt{\det(1-\theta \hF)}  \hF \frac{1}{1-\theta \hF} \;       
W(x,\cC_k) \right] \ast e^{ik \cdot x} \\
\label{q4}
Q_{4} (k) & = \frac{1}{2} \Tr \int d^{p+1} x \; L_{\ast} \left[
\sqrt{\det(1-\theta \hF)}  \left(\hF \frac{1}{1-\theta \hF}\right) 
\wedge     \left(\hF \frac{1}{1-\theta \hF}\right)
W(x,\cC_k) \right] \ast e^{ik \cdot x}
\end{align}
Then the \RR\ couplings can be written as
\begin{equation} \label{nonc}
\swz = \int \left[Q_0 D^{(p+1)} + Q_2  D^{(p-1)}
+  Q_4  D^{(p-3)} + \cdots \right] \ . 
\end{equation} 

The gauge invariance under $D \rightarrow D + d\Lambda$ requires 
$Q (x)$ to be closed, a statement we have  only checked for 
some special cases. Nevertheless, here we shall assume it and explore 
its implications. For $Q_0$ to be closed it must be a constant,
and it can be checked that $Q_0 (x) = N$, which implies the following  
interesting identity
\begin{equation} \label{unit}
\Tr \int d^{p+1} x \; L_{\ast} \left[
\sqrt{\det(1-\theta \hF)}\;  W(x,\cC_k) \right] \ast e^{ik \cdot x}
= N (2 \pi)^{p+1} \delta^{(p+1)} (k)  \ .
\end{equation}
That $Q_2$ is closed  implies that locally there is a one form $A$ so that 
$d A = Q_2$, i.e. 
\begin{equation} \label{swm}
(d A) (k) = \Tr \int d^{p+1} x \; L_{\ast} \left[
\sqrt{\det(1-\theta \hF)}  \hF \frac{1}{1-\theta \hF} \;       
W(x,\cC_k) \right] \ast e^{ik \cdot x}
\end{equation} 
For a $U(1)$ gauge group, equation~\eqref{swm} is precisely the
Seiberg-Witten map
conjectured in~\cite{liu}. Thus the consistency of the couplings
essentially confirms the conjecture of~\cite{liu}. While the  conjecture 
has a natural generalization to the $U(N)$ case,~\eqref{swm} appears
to only confirm the $U(1)$ part of it.

Comparison between the \RR\ couplings in different descriptions  can give
a better understanding of the Seiberg-Witten map~\eqref{swm},
and also constrain $\apr$ corrections. In the commutative description,
it has been argued in~\cite{nick} that there is no $\apr$ 
correction for the couplings of $D^{p+1}$ and $D^{p-1}$ in~\eqref{dbc}.
That $Q_0 = N$ and $Q_2$ precisely gives the Seiberg-Witten
map for the gauge fields, is in accord with the results of~\cite{nick} and 
also suggests that there are no $\alpha'$ corrections to~\eqref{nonc}
in the $\Phi=0$ description either, a conclusion
which is otherwise hard to obtain
from amplitude calculations. For the couplings of $D^{(p-3)}$, there 
are higher order $\alpha'$ corrections to both descriptions; some corrections 
in the commutative description were found in~\cite{nick} and in the $\Phi=0$
description they can be seen from the disk amplitude with two open string 
insertions.  Thus in this case we cannot conclude that $Q_4 = F \wedge F$ 
and it should be interesting to constrain the higher order 
corrections on both side by using  the explicit Seiberg-Witten map~\eqref{swm}.
What we can conclude is that since $\int Q_4$ gives the topological 
charge for the D$(p-4)$-brane (see section~\ref{ktheory}) 
in the   $\Phi=0$ description
and $\Tr \int F \wedge F$ gives the same charge in the commutative 
description, they must be identified, i.e. we must have 
\begin{equation} \label{topo}
Q_4 (k=0) = \frac{1}{2} \Tr \int F \wedge F
\end{equation}
where $Q_4$ is given by~\eqref{q4}. Similar statements can also be made for 
$Q_6$ and $F^3$ and so on.  
 
We now know the leading order couplings to the \RR\ potentials
for all three special cases listed in the Introduction. One naturally 
wonders about the story for other values of $\theta$. We would like to 
argue that the action~\eqref{rrf} and its finite momentum counterpart 
apply to all $\theta$-descriptions. More precisely,
we claim that, if we fix the closed string background $g,B$, then for any 
choice of $\theta$, the leading order couplings (in terms of
an $\apr$ expansion) 
to \RR\ potentials are given by~\eqref{rrf} and its finite momentum 
counterpart.
A heuristic argument is as follows.
\begin{enumerate}

\item  The couplings to zero momentum \RR\ potentials should give topological 
charges on the branes for any $\theta$ description. \label{zero}

\item Our discussions of  the 
topological charges
in noncommutative gauge theories in section~\ref{ktheory} and the 
identities~\eqref{unit},~\eqref{topo} hold for all 
values of $\theta$. \label{one}

\item The finite momentum couplings are obtained by smearing
the zero momentum couplings to an open Wilson line using the 
$L_{\ast}$ prescription (see section~\ref{fmc})%
\footnote{Of course, we have only checked the $L_{\ast}$ prescription 
in the $\Phi=0$ description from the amplitudes. However, that it 
should be true for all descriptions is supported by the 
$\theta$ independence of the Seiberg-Witten map~\eqref{swm}.}. \label{two}

\end{enumerate}
Thus from the above statements we conclude the zero momentum 
couplings for all $\Phi$ should be given by~\eqref{grr} and at finite momentum
by~\eqref{gfmc} and~\eqref{gfmc1}. 
Another immediate consistency check is 
that by taking $\theta = 0$ and $\theta = \frac{1}{B}$ along with a change 
of gauge field variables $\hA$ appropriate for each $\theta$ in~\eqref{rrf} 
we indeed recover the results for the commutative and the Matrix descriptions.

In the above we have concentrated on the part of the action 
involving the  gauge fields. Completely parallel statements regarding 
the Seiberg-Witten map and $\apr$ corrections can be made 
for the tranverse scalar fields
by starting with D$9$-branes and doing dimensional reduction or T-duality.
It would be interesting to work them out explicitly.

\skipthis{
to get at least 
topological parts of the couplings for any $\theta$, since all these identities
hold for all values of $\theta$
(or 
equivalently~\eqref{asXX},~\eqref{trr}). An immediate consistency check is 
that by taking $\theta = 0$ and $\theta = \frac{1}{B}$ along with a change 
of gauge field varaibles $\hA$ appropriate for each $\theta$, 
we indeed recover the results for the commutative and the Matrix descriptions. 
Other evidence for this claim is to start with the commutative 
description result~\eqref{dbc} and then use 
the identities like~\eqref{unit},~\eqref{swm},~\eqref{topo} to get at least 
topological parts of the couplings for any $\theta$, since all these identities
hold for all values of $\theta$. Then the finite momentum part is bulit by the 
procedure using the results of section~\ref{fmc}. 
}

\section{Derivative-driven dielectric effects} \label{transverse}

In this section we shall be interested in 
the dynamics of the transverse scalar fields. 

When $B=0$,
an important aspect of the story in the commutative description
is the presence of a 
factor $e^{-i \co{\phi}\co{\phi}}$ in~\eqref{dbc}, which couples D$p$-branes 
to \RR\ potentials of rank 
higher  than their dimensions(e.g. to $C^{(p+3)}$). While in a noncompact 
transverse space, it is not possible for finite number of D$p$-branes to carry
a net D$(p+2)$ brane charge for obvious topological reasons, they can carry 
dipole or multipole moments, when put in a 
background \RR\ field strength $F^{(p+4)}$~\cite{myers}. 
These dielectic effects, 
which turn the Chan-Paton degrees of freedom into  a spacetime fuzzy 
geometry, require the number of branes $N > 1$. 

When we turn on a constant $B$ field, while we expect the above story to 
remain essentially the same, it is now possible to generate the dielectric 
effects from a single brane. The reason is that there are possible 
higher order $\apr$ corrections like%
\footnote{We would like to thank C. Hofman for discussions regarding 
this point.}%
\begin{equation} \label{deri}
\int \apr^3 B_{\mu \nu} B_{\lambda \rho} B_{\tau \sigma} \;
\left(\p^{\mu} \p^{\lambda}  \phi^i\right)
\left(\p^{\rho} \p^{\tau}  \phi^j\right)
\left(\p^{\nu} \p^{\sigma}  \phi^k\right)
F^{(p+4)}_{ijk} 
\end{equation} 
to~\eqref{dbc} when $B \neq 0$. Thus it is possible to have dielectric 
effects  purely driven by the nontrivial profiles of transverse scalar 
fields in the world-volume. 

In this section we would like to analyze such derivative-driven dielectric 
effects using the noncommutative description~\eqref{rrf} and its finite 
momentum counterpart. The advantage is that  terms like~\eqref{deri} 
are already present in the leading terms~\eqref{rrf} in the noncommutative 
description,%
\footnote{Recall that the effective action in one description is 
a resummation of that in another description including all orders in 
$\apr$ corrections.}
thus enabling a more systematic analysis. Also in terms of~\eqref{asXX}
a single D$p$-brane with a constant $B$-field may be considered as 
a collection of an infinite number of D$0$-branes. Thus that we shall 
have the dielectric effects for a single brane is not surprising after all. 

It is important to observe that in~\eqref{rrf} in addition to the 
factor  $e^{-i \co{\hp}\co{\hp}}$, the contraction $e^{\co{\theta}}$ 
acting on the terms in the pull-back also generate  couplings 
to \RR\ forms of higher rank. This is crucial to the absence of the 
coupling to $C_{ij}^{(p+3)}$ of the following form:
\begin{equation} \label{charge}
\int_{{\cal B}_p} \; f (\hA)  \hp^i\hp^j C_{ij}^{(p+3)}
\end{equation}
where $C_{ij}^{(p+3)}$  should be considered as 
a world-volume (denoted by ${\cal B}_p$) $(p+1)$-form and 
$f$ is fuction of $\hA$. In the above equation  
the precise product structure, ordering or possible 
derivatives on $\hp^i$ are not important. 
The presence of such a coupling would 
imply the possibility of generating a net D$(p+2)$-brane charge. 
This is a very nontrivial statement, implying all terms of the form 
$\hA^n \hp \hp$ (for some integer $n$) in the finite momentum version 
of~\eqref{rrf} should cancel. We have checked the cancellations for
the two lowest orders,
$\hp \hp$ and  $\hA \hp \hp$. For example, at lowest order this
is guaranteed by the following identity~\cite{mehen,liu}:
\begin{equation} \label{id}
\Tr \left(i \com{\hp^{i}}{\hp^{j}}_{\ast} 
+ \half \theta^{\sigma\tau} \p_\sigma X^i \ast_2 \p_\tau X^j \right)
= 0
\end{equation}
which is a special case of the  recursion 
relation~\cite{liu} (see also~\cite{kiem})
\begin{equation} \label{recur}
\theta^{\mu\nu} \p_\nu \ast_n[f_1,\dots,f_{n-1}, \p_\mu g] = i
\sum_{j=1}^{n-1} \ast_{n-1} [f_1,\dots,f_j\ast g-g\ast f_j, \dots,
f_{n-1}].
\end{equation}

The dielectic effects appear when we look at the terms cubic in $\hp$.
For a constant \RR\ field strength $F^{(p+4)}_{ijk} 
= 2 f \ep_{ijk} \ep_{(p+1)} $ we find a coupling 
(for simplicity, we consider the $U(1)$ case)
\begin{equation} \label{preeom}
\frac{i}{3} \int \left( \frac{i}{2} \com{\hp^i}{\hp^j} + 
\frac{1}{4} \theta^{\mu \nu} \anti{\p_{\mu} \hp^i}{\p_{\nu} \hp^j}
\right) \hp^k \; F^{(p+4)}_{ijk}
\end{equation}
where $\anti{\cdot}{\cdot}$ is an anticommutator.
We shall assume that $\theta$ is large, so that we can ignore the 
quadratic kinetic term $\del^2 \phi^i$ from the Born-Infeld action.
Combining with the $\com{\hp^i}{\hp^j}^2$ terms in the Born-Infeld 
action, we get an equation of motion 
\begin{equation} \label{eom}
\com{\com{\hp_i}{\hp_j}}{\hp^j} = i f \ep_{ijk} \left( \com{\hp^j}{\hp^k} - 
\frac{i}{2} \theta^{\mu \nu} \anti{\p_{\mu} \hp^j}{\p_{\nu} \hp^k}
\right)
\end{equation}
All products in equations~\eqref{preeom} and~\eqref{eom} are
$\ast$-products and $i,j,\cdots$ indices
are raised and lowered by the metric $g_{ij}$. All but
the second term on the right hand side of the equation
is  simply the generalization of  equations for the commutative 
nonabelian case to the $\ast$-product algebra. The new term, which arises 
from the contraction $\co{\theta}$ with the quadratic terms in the 
pull-back, has the intriguing stucture of a Poisson bracket with 
respect to $\theta$ superposed with an $SU(2)$ algebra in terms 
of the $\ast$-product. The above equations should give a derivative 
driven dielectric effect. It would be very interesting to analyze in more
detail the solutions of~\eqref{eom}. However,
an exact solution appears hard to come by. We hope to return to this
question in the future.

\skipthis{
On the noncommutative plane, and in the usual Hilbert space formalism
of noncommutative geometry
(see e.g.~\cite{gms,gn}
\comment{and which Seiberg paper?}), equation~\eqref{eom} has the simple and
suggestive form
\begin{equation}
\com{\hp^j}{\com{\hp^i}{\hp^j}} = 
i f \epsilon_{ijk} \com{\hp^j}{\left(\anti{a^\dagger}{\com{a}{\hp^k}}
  - \hp^k \right)}
-i \frac{f}{2} \epsilon_{ijk} \anti{a^\dagger}{\com{a}{\com{\hp^i}{\hp^j}}},
\end{equation}
where $a = \frac{x^1+ix^2}{\sqrt{2\theta}}$ and
$a^\dagger = \frac{x^1-ix^2}{\sqrt{2\theta}}$.
However,
an exact solution appears hard to come by. We hope to return to this
question in the future.
}

\skipthis{

Na\"{\i}vely, it would appear that the factor $e^{-i \co{X}\co{X}}$
in~\eqref{rrc} leads to Myers-type terms.  In fact, as we will
discuss, this is true, but the situation is somewhat more subtle.  In
particular, in ref.~\cite{myers}, it was emphasized that there are no
terms in the action that are quadratic in the transverse scalars.
This is a purely algebraic statement: $\Tr\com{X}{X}=0$ for $X\in
U(N)$.  Here, of course, there is no such restriction; nevertheless,
it appears that the quadratic term does not appear.  Demonstrating
this is quite nontrivial, and we have only shown it to low orders in
powers of the gauge field.  The demonstration relies on the recursion
relation~\cite{liu}\footnote{This corrects an obvious typo in
equation~(2.19) of~\cite{liu}.}
\begin{equation} \label{recur}
\theta^{\mu\nu} \p_\nu \ast_n[f_1,\dots,f_{n-1}, \p_\mu g] = i
\sum_{j=1}^{n-1} \ast_{n-1} [f_1,\dots,f_j\ast g-g\ast f_j, \dots,
f_{n-1}].
\end{equation}
Using~\eqref{recur}, one finds, remarkably, that the terms for which
$\co{\theta}$ acts on the quadratic terms in the  pullback, precisely
cancels the quadratic terms coming from $e^{-i \co{X}\co{X}}$.  
To zeroth order in the gauge field, this is
the statement that
\begin{equation} \label{id}
\Tr \left(i X^{[i} \ast X^{j]} 
+ \half \theta^{\sigma\tau} \ast_2[\p_\sigma X^i,\p_\tau X^j]\right)
= 0,
\end{equation}
which follows from~\eqref{recur}, or equivalently $\ast_2 =
\frac{\sin\frac{k_1\times k_2}{2}}{\frac{k_1\times k_2}{2}}$,
equation~\eqref{cfmyers} and the vanishing of the trace of a $U(N)$ commutator.
One might have expected gauge invariance to imply that the statement
is then true to all orders in the gauge field; however, it turns out
that already at linear order, the incorporation of the factor $e^F$---which is
not required by gauge invariance---is required for the continued
vanishing of the terms that are quadratic in $X$.

This
surprising result is made perhaps less surprising by noting that any
quadratic term in the action would lead to a linear term in the
equations of motion, thus preventing the usual $\com{X}{X}=0$ solution.
Actually, when the equation of motion is computed, it turns out that
the trivial
solution is not $\com{X}{X}_\ast=0$, but $\theta^{\mu\nu}\p_\mu X^i
\p_\nu X^j = 0$; i.e.\ the vanishing of the Poisson bracket of the $X$s.

}

\section{Discussions} \label{conclusion}

In this paper we have investigated the \RR\ couplings of 
noncommutative D-branes. We argued that the leading couplings we found 
for the $\Phi = 0$ description apply to all $\Phi$-descriptions. The zero 
momentum couplings are topological, involving the Elliott formula, and thus 
are universal for all descriptions. The finite momentum couplings are 
determined partly by gauge invariance and the universality of the 
Seiberg-Witten map. 

Our results should have a variety of applications, in addition to
those we have
already discussed in the paper.
For example, from~\eqref{grr} and~\eqref{gfmc}, 
turning the argument of~\cite{ghm} backwards, we may deduce
the presence of chiral fermion zero modes and gain insight 
into e.g.~the index theorem in the noncommuative geometry. 
For example consider a Weyl fermion on a $2p$-dimensional 
noncommutative manifold with a Yang-Mills connection $\hA$. The anomalous 
variation of the action $\log Z(\hA)$ then could be given by
\begin{equation}
\delta \log Z(\hA) \sim \int \bigl[ e^{-\co{\theta}} \ch (\hF) 
\bigr]^{(1)}
\end{equation}
which can be considered as the noncommutative generalization of the famous 
descent formula. Such a formula can also be postulated based on our discussion
of the Seiberg-Witten map in section~\ref{phi} (see e.g~\eqref{topo}). 

It would be interesting to generalize the present discussion to non-BPS
branes and brane-antibranes. In these cases the formula~\eqref{dbc} still 
holds (let us for simplicity set the transverse scalars to zero); however 
one must replace the connection $A$ and curvature $F$ there by  
the so-called superconnection ${\cal A}$ and supercurvature 
${\cal F}$ (see e.g. ~\cite{andy,per,tera,schwarzw}).  
From the results of the present paper, it is then tempting to speculate 
that in the noncommutative case, one simply 
replaces the Chern character for the superconnection by a ``super-Elliott''
formula
\begin{equation}
\int \, C \; e^{-\co{\theta \tau}} \; e^{{\cal \hF}} 
\end{equation}
where $\tau$ takes value $1$ in the non-BPS case and $\tau= \sigma_3$ (Pauli
matrix) in the D$\bar{\text{D}}$ case, since in the D$\bar{\text{D}}$
case brane and anti-brane have opposite noncommutative parameter due to
orientation. It would be interesting to check explicitly whether the above
simple extrapolation is realized.

\skipthis{
\section{Appendix}

\skipthis{
Recalling our abusive equation~\eqref{altPb}, and noting the identity
\begin{equation} \label{moveco}
\int \left(e^{-\co{\theta}} \chi\right) \omega 
= \int \chi e^{-\co{\theta}} \omega,
\end{equation}
which follows from
\begin{equation} \label{itow}
e^{-\co{\theta}} \omega = \Pf(\theta) \star \left( \left(\star
e^{-\theta^{-1}} \right) (\star \omega) \right),
\end{equation}
and%
\skipthis{\footnote{For an
even-dimensional Euclidean space, the sign is correct for
even-dimensional forms.  Since the space in question is
that spanned by the noncommutative directions---with time commutative
for even $p$---this is always the case for our expressions.}}%
\begin{equation} \label{nostar}
\int (\star \omega) (\star \chi) = \int \chi \omega
\end{equation}
where $\star$ is the Hodge dual%
\footnote{The metric dependence of the Hodge dual drops out in these
equations.}
in the {\em noncommutative}%
\footnote{That is, for odd $p$, we take the Hodge dual in the brane,
whereas for even $p$, we take the Hodge dual along the spatial
directions of the brane. \label{ft:hd}}
directions and
$\Pf$ is the Pfaffian (using the conventions of~\cite{naka}),
we can rewrite
equation~\eqref{rrc} as
\begin{equation} \label{rrc2}
L_\ast \int W_q[\cC(q)] e^F e^{D\co{\hp}} e^{-\co{\tilde\theta}}
  e^{-i \co{\hp} \co{\hp}}  C(-q) e^B.
\end{equation}
As we will show momentarily, 
if we repeatedly use~\eqref{itow} and~\eqref{nostar}, 
then we can turn
this into
\begin{equation} \label{asXX}
\frac{1}{\Pf(\theta)}
L_\ast \int W_q[\cC(q)] \star \left(e^{-i \co{\com{\cX}{\cX}}} C(-q)
e^{B-\frac{1}{\theta}}\right),
\end{equation}
where $\cX^\mu = x^\mu + \theta^{\mu\nu} A_{\nu}$ and $\cX^i=X^i$.
For $p=0$, we can ignore the Hodge dual (as per footnote~\ref{ft:hd}),
and then~\eqref{asXX} is a concise way of writing the \WZ\ couplings
very recently given in~\cite{oo}.

To demonstrate~\eqref{asXX}, we use the Campbell-Baker-Hausdorff
formula to find
\begin{equation} \label{cbh}
e^{D\co{X}} e^{-\co{\theta}} = e^{-\co{\theta}} e^{\frac{1}{\theta}}
  e^{\co{\theta DX}} e^{-\frac{1}{\theta}}.
\end{equation}
}
So,
\begin{alignat}{2}
&&& L_\ast \int W_q[\cC(q)] \left(e^{-\co{\tilde\theta}} e^F \Pb\right)
  e^{-i \co{X}\ast\co{X}}  C(-q) e^B \nonumber \\
&\text{\small{[via~\eqref{moveco}]}}
&& = L_\ast \int W_q[\cC(q)] e^F e^{D\co{X}} e^{-\co{\tilde\theta}} 
  e^{-i \co{X}\ast\co{X}}  C(-q) e^B \\
&\text{\small[via~\eqref{nostar},~\eqref{cbh}]}
&&= L_\ast \int W_q[\cC(q)] \left(\star e^F\right)
  \left( \star  e^{-\co{\theta}} e^{\frac{1}{\theta}}
  e^{\co{\theta DX}} e^{-\frac{1}{\theta}}
  e^{-i \co{X}\ast\co{X}}  C(-q) e^B \right)  \\
&\text{\small[via~\eqref{itow}]} 
&&= L_\ast \int W_q[\cC(q)] \Pf(-F) \star \left[ e^\co{F^{-1}-\theta}
  e^{\frac{1}{\theta}} e^{\co{\theta DX}} 
  e^{-i \co{X}\ast\co{X}}  C(-q) e^{B-\frac{1}{\theta}} \right] \\
\label{mukhi}
&\text{\small[via~\eqref{itow},~\eqref{nostar}]} 
&&= L_\ast \int W_q[\cC(q)] \frac{\Pf(\theta-\theta F \theta)}{\Pf(\theta)}
   e^{\frac{1}{\theta-\theta F \theta}}
   e^{\co{\theta DX}} 
  e^{-i \co{X}\ast\co{X}}  C(-q) e^{B-\frac{1}{\theta}} \\
&&& = \frac{1}{\Pf(\theta)}
L_\ast \int W_q[\cC(q)] \star \left(e^{-i \co{\com{\cX}{\cX}}} C(-q)\right)
e^{B-\frac{1}{\theta}}.
\end{alignat}
}

\acknowledgments

We are grateful for useful conversations with  M. Douglas, J. Harvey, 
F. Larsen,  L.~Motl, S. Sethi, and especially C.~Hofman, E. Martinec and 
G.~Moore for important discussions. We would also like to thank C. Hofman
for a careful reading of the manuscript.
This work was supported by DOE grant \hbox{\#DE-FG02-96ER40559}.  
J.M. was also supported by an NSERC PDF Fellowship.  J.M. thanks the ITP 
in Santa Barbara, where this work was completed with a little help from Grant 
No.\ PHY99-07949 of the National Science Foundation. H.L. also thanks 
the Chicago
theory group for hospitality.

\end{document}